\documentclass[sigconf, none, 10pt]{acmart}
\usepackage{framed}
\usepackage{url}
\usepackage{graphicx}
\usepackage{flushend}
\usepackage{cleveref}
\usepackage{amssymb}
\usepackage{amsmath}
\usepackage{xcolor}
\usepackage{soul}
\usepackage{algpseudocode}
\usepackage{algorithm}
\usepackage{xspace}
\usepackage{tabularx}
\usepackage{multirow}
\usepackage{balance}
\usepackage{subcaption}
\usepackage{booktabs} 
\usepackage{booktabs}
\newenvironment{tab}[1]
{
\let\oldarraystretch=\arraystretch
\renewcommand{\arraystretch}{1.1} 
\begin{tabular}{@{}#1@{}}
\toprule
}
{\bottomrule
\end{tabular}
\renewcommand{\arraystretch}{\oldarraystretch}
}

\PassOptionsToPackage{hyphens}{url}
\expandafter\def\expandafter\UrlBreaks\expandafter{\UrlBreaks
	\do\a\do\b\do\c\do\d\do\e\do\f\do\g\do\h\do\i\do\j%
	\do\k\do\l\do\m\do\n\do\o\do\p\do\q\do\r\do\s\do\t%
	\do\u\do\v\do\w\do\x\do\y\do\z\do\A\do\B\do\C\do\D%
	\do\E\do\F\do\G\do\H\do\I\do\J\do\K\do\L\do\M\do\N%
	\do\O\do\P\do\Q\do\R\do\S\do\T\do\U\do\V\do\W\do\X%
	\do\Y\do\Z}

\newcount\Comments  \Comments=1  
\newcommand{\kibitz}[2]{\ifnum\Comments=1\textcolor{#1}{#2}\fi}

\providecommand{\vs}{vs. }
\providecommand{\ie}{\emph{i.e.,} }
\providecommand{\eg}{\emph{e.g.,} }

\providecommand{\etal}{\emph{et al.\xspace}}

\providecommand{\myparab}[1]{\smallskip\noindent\textbf{#1} }

\newcommand{\sectionref}[1]{$\S$\ref{#1}}

\newcommand{\squishenum}{\begin{enumerate}{}{\setlength{\itemsep}{0pt}\setlength{\parsep}{0pt}\setlength{\topsep}{3pt}\setlength{\partopsep}{0pt}\setlength{\leftmargin}{1.5em}\setlength{\labelwidth}{1em}\setlength{\labelsep}{0.5em}}}
\newcommand{\squishlist}{\begin{list}{$\bullet$}{\setlength{\itemsep}{0pt}\setlength{\parsep}{3pt}\setlength{\topsep}{3pt}\setlength{\partopsep}{0pt}\setlength{\leftmargin}{1.5em}\setlength{\labelwidth}{1em}\setlength{\labelsep}{0.5em}}}
\newcommand{\squishlisttwo}{\begin{list}{$\bullet$}{\setlength{\itemsep}{0pt}\setlength{\parsep}{0pt}\setlength{\topsep}{0pt}\setlength{\partopsep}{0pt}\setlength{\leftmargin}{2em}\setlength{\labelwidth}{1.5em}\setlength{\labelsep}{0.5em}}}
\newcommand{\squishend}{\end{list}}
\newcommand{\squishenumend}{\end{enumerate}}

\setcopyright{none}
\begin{document}
\title{A Churn for the Better}
\subtitle{Localizing Censorship using Network-level Path Churn and
Network Tomography}

\author{Shinyoung Cho $^1$, Rishab Nithyanand $^1$, Abbas Razaghpanah $^1$, Phillipa Gill $^2$}
\affiliation{$^1$ Stony Brook University, $^2$ University of Massachusetts, Amherst}

\renewcommand{\shortauthors}{S. Cho et al.}

\begin{abstract}

Recent years have seen the Internet become a key vehicle for citizens
around the globe to express political opinions and organize protests. This fact
has not gone unnoticed, with countries around the world repurposing network
management tools (\eg URL filtering products) and protocols (\eg BGP, DNS) for
censorship. However, repurposing these products can have unintended
\emph{international} impact, which we refer to as ``censorship leakage''. While
there have been anecdotal reports of censorship leakage, there has yet to be a
systematic study of censorship leakage at a global scale.


In this paper, we combine a global censorship measurement platform (ICLab)
with a general-purpose technique -- \emph{boolean network tomography} -- to
identify which AS on a network path is performing censorship. At a high-level,
our approach exploits BGP churn to narrow down the set of potential censoring
ASes by over 95\%. We exactly identify 65 censoring ASes and find that the
anomalies introduced by 24 of the 65 censoring ASes have an impact on users
located in regions outside the jurisdiction of the censoring AS, resulting in
the \emph{leaking} of regional censorship policies.

\end{abstract}

%
%

\maketitle

\section{Introduction}\label{sec:introduction}
The Internet is now regarded as part of the critical infrastructure, with
citizens relying on it for dissemination of information and organizing political
action. Consequently, governments and network-level entities -- \eg Autonomous
Systems (ASes) -- are implementing forms of censorship to restrict access to
specific content, in many cases, repurposing existing network management tools
\cite{Dalek2013} and protocols (\eg DNS~\cite{Anonymous-FOCI14, Levis2012}, 
BGP~\cite{HIJACKS-pakistan}) to filter Internet content. The past decade has seen
numerous instances, where this repurposing of protocols has had unintended
international impact, which we refer to as ``censorship leakage.'' These have
included the 2008 hijack of YouTube traffic by Pakistan
Telecom~\cite{HIJACKS-pakistan}, and cases of DNS root-servers located in China
impacting international users~\cite{DNS-china}. While specific instances of
censorship leakage have been a boon for researchers (in particular those
studying China)~\cite{Anonymous-FOCI14, Xu2011, Levis2012, Crandall2007,
Weaver2009}, we lack a broader understanding of the prevalence of this
phenomenon. 
Part of the challenge of understanding censorship leakage on a global scale is
developing a technique that is able to identify censorship leakage in general
\vs looking for specific instances (\eg injected RST or DNS packets from
China~\cite{Anonymous-FOCI14, Weaver2009}). This challenge is compounded by a
general lack of vantage points that are available for measuring censorship on an
ongoing basis. We address these challenges by combining  the ICLab measurement
platform with the idea of \emph{boolean network tomography} \cite{Vardi1996}.
ICLab is a platform of $\sim$1,000 globally distributed vantage point that has
been performing measurements of censorship on an ongoing basis since November
2015 (more details in~\sectionref{sec:background:datasets}).

Our intuition is that we can observe multiple tests from a given  vantage point
to a given destination and that, if there is sufficient path churn between the
vantage point and destination, we can create a set of boolean constraints where
the constraint is true if censorship is observed, and false otherwise. In the
case where censorship is observed, it must be the case that at least one
autonomous system (AS) on the path is performing censorship. We can then input
these constraints into an off-the-shelf SAT solver to identify the AS performing
censorship. In this study, we demonstrate the applicability of this intuition by
answering the following key questions: (1) {Is there enough path churn observed
in our measurements to create a solvable set of constraints?} We need to
validate that we have enough variability in paths, especially in the cases where
we observe censorship, to create a solvable set of constraints to narrow down
the set of potential censors. (2) {Will our constraints generate a small set of
potential censoring ASes?} We want to make sure that the set of potential
censoring ASes is not intractably large, making it impossible to exactly
identify ASes responsible for implementing censorship. While answering these
questions, we make the following contributions:

\myparab{Problem reformulation. }We demonstrate how measurements gathered by the
ICLab platform can be used to formulate a boolean network tomography problem
solvable by off-the-shelf SAT solvers. Our approach carries over to other
measurement databases such as those generated by the OONI \cite{Filasto2012} and
the M-Lab \cite{MLab} platforms.

\myparab{Measuring and exploiting network-level churn.} We show that the
instability of network-level paths can act as a
substitute for strategically placed internal monitors. Specifically, we
show that 25\%, 30\%, 28\%, and 67\% of paths between ICLab vantage points and
web servers are observed to change over periods of one day, week, month, and
year. These changes are found to significantly improve the solvability of our
constructed SAT problems. 

\myparab{Identifying censors and censorship leakage.} We empirically demonstrate
that our approach allows us to reduce the size of the set of potential censoring
ASes by over 95\%, on average. Further, we exactly identify 65 censoring ASes
located in 30 different countries. Our study also identifies \emph{leakage} of
censorship policies -- \ie cases where censoring ASes blocked access to content
even for users outside their country of operation. Specifically, we find that 32
and 24 of the censoring ASes leak censorship to other ASes and countries,
respectively.

%


\section{Background \& Related work}\label{sec:background:related}
\myparab{Censorship measurement. }
Much of previous work has focused on understanding how censorship is performed
by network-level entities. Studies have shown that censors may restrict access
to content by injecting incorrect DNS replies \cite{Nabi-FOCI13,
Anonymous-FOCI14}, sending TCP reset packets spuriously \cite{Khattak-FOCI13,
Aryan-FOCI13}, using off-the-shelf filtering and blocking tools
\cite{Dalek-IMC13}, or throttling connections to censored content
\cite{Aryan-FOCI13}. Our study builds off of related work on large-scale
longitudinal censorship measurement systems. Specifically, we use data collected
by the ICLab platform~\cite{Abbas2016} to detect censorship and generate
constraints. {Conceptually, our techniques could be applied to other
platforms such as OONI~\cite{Filasto2012} as well.}

\myparab{{Fault localization.} }
Other studies have approached the problem of network failure localization with
different perspectives. Lifeguard \cite{Katz-Bassett2012} relies on historical
control-plane measurements and active probing to automatically identify and
route around network failures via crafted BGP messages. 
{Feamster \etal \cite{Feamster2003} measure the effectiveness of reactive
routing around node failures. Their approach localizes failures in real-time by
analyzing results of active probes, including pings and traceroutes, between
vantage points. While other work in the area has focused on identifying the root
cause of  path changes on the Internet \cite{Javed2013, Feldmann2004,
Teixeira2004, Wu2005, Pei2005}, we focus on localizing faults and errors that do
trigger path changes. 
 
\myparab{Boolean network tomography. }
Network tomography \cite{Vardi1996}  typically involves using end-to-end
measurements and a set of monitors within the network to uncover hidden node
values (which in the case of boolean network tomography, may only take the
values True or False). Monitors are used to ensure that appropriate
end-to-end measurements may be performed to unveil specific node
characteristics. We use the data gathered by the ICLab platform as
end-to-end measurements from which we identify nodes (ASes) implementing
specific types of censorship. Unlike typical boolean network tomography
problems, our study is limited by the absence of strategically located monitors
from which end-to-end measurements can be gathered. However, we show that due to
the churn of network-level paths, we are still able to use
boolean network tomography to identify censoring ASes.

Several studies have focused on the problem of error localization through
boolean network tomography. Ma \etal \cite{Ma2017} focus on identifying the
conditions, monitor locations, and probing mechanisms that are necessary for
fault localization through boolean network tomography. Dhamdhere \etal
\cite{Dhamdhere2007} use a boolean network tomography approach in conjunction
with ``troubleshooting sensors'' (monitors) located within the network to
identify misconfigured routers responsible for network failures. Other work has
applied boolean network tomography to identify areas of heavy congestion
\cite{Tsang-2003} and packet loss \cite{Coates-2000, Coates-2002}. In this
paper, we use boolean network tomography combined with the
network-level path churn in routing protocols (as a substitute
for specific monitors) to identify ASes that introduce censorship
related anomalies.

\subsection{The ICLab Dataset}\label{sec:background:datasets}

%
%

We rely on data gathered by the ICLab censorship measurement platform
\cite{Abbas2016} as a source for end-to-end measurements. The ICLab platform
repetitively performs a variety of measurements between a set of over 1K
globally distributed vantage points and web-servers hosting regionally sensitive
content. The platform aims to (1) identify content being censored, (2)
understand how censorship is implemented, and (3) record changes in censorship
policies over time. Specifically, ICLab identifies the following anomalies as
indicative of potential censorship:

\myparab{DNS anomalies. }DNS anomalies occur when a censor injects  DNS
responses to queries issued by a client. The idea being that if the censor is
closer to the  client, their injected packet will arrive at the client before
the response from the DNS resolver. DNS injection can be detected by observing
two response packets for the same DNS request.  The ICLab platform identifies
DNS injection by making DNS queries for domain names using both, Google's DNS
resolver (\texttt{8.8.8.8}) and the default resolver of the vantage point, and
then observing the number of response packets received, in each case. If a
second DNS response packet is received within two seconds of the first, an
anomaly is reported.

\myparab{SEQNO and TTL anomalies.} A packet injector will often have attributes
that differ from the legitimate server for a connection. ICLab platform issues
HTTP GET requests and records all responses (while following redirects). The
platform then analyzes raw packet captures to identify anomalies. Specifically,
we compare the IP TTL header on the SYNACK packet of the connection with
subsequent packets. This relies on the assumption that a censor will not be fast
enough to act prior to the SYNACK being sent by the server.  Further, injected
packets will often not be able to perfectly mimic the TCP state of the
server~\cite{Weaver2009}. We look for cases where there are overlapping sequence
numbers between packets or gaps in sequence numbers. These sequence number
anomalies, especially when combined with packets having the RST flag set (to
close the connection) are likely indicators of censorship.
%

\myparab{Block pages. }Finally, the platform analyzes the responses received to
identify blockpages that are returned by a censor. This is done by performing
regular expression matching with known examples of blockpages (provided by the
OONI project~\cite{ooni-site}) and by comparing responses with those obtained
from censor-free vantage points within the United States. In the latter case, we
employ techniques developed by Jones~\emph{et al.} to identify block
pages~\cite{Jones2014}

\myparab{Network paths.} In addition to gathering the above data, the platform
also records traceroutes from vantage points to the corresponding destinations
of each test. 

In total, we utilize 4.9M measurements (with 39K total identified
anomalies) from the platform} between vantage points located in 539 different
ASes (in 219 countries) and 774 URLs. A summary of the ICLab data that we use in
our work is summarized in \Cref{table:data-summary}.

\begin{table}[t]
\centering
\small
\begin{tabular}{ll}
\hline
Period             & 2016-05 $\sim$ 2017-05 \\
\hline
Unique URLs        & 774                    \\
AS Vantage Points  & 539                    \\
Destination ASes   & 620                    \\
Countries          & 219                    \\
\hline
Measurements       & 4.9M                   \\
 -- w/DNS anomalies & 2.3K	(0.05\%)   \\
 -- w/SEQNO anomalies & 9.8K	(0.20\%)   \\
 -- w/TTL anomalies & 17K	(0.35\%)   \\
 -- w/RESET anomalies & 8.4K	(0.17\%)   \\
 -- w/Blockpages   & 1.5K	(0.03\%)   \\ 
\hline
\end{tabular}
\caption{ICLab dataset characteristics.}
\vspace{-.2in}
\label{table:data-summary}
\end{table}

\myparab{Ethical considerations and limitations. }
To mitigate risk, the vast majority of ICLab's vantage points are obtained via
commercial VPN providers (many of which are located in ASes classified as
content ASes by CAIDA\cite{CAIDA-ASClass}). This allows us to obtain widespread continuous
measurements, without putting users in specific regions at risk. A potential
limitation of this decision, is the inability to observe the same filtering as
ASes providing residential connections.

In collaboration with the Citizen Lab, we have worked to deploy a handful of
Raspberry Pi nodes running the measurement software. Prior to deploying a node,
we discuss with the volunteer about the potential risks and they are further
presented with a form that summarizes risks for their given country based on
existing metrics (\eg Freedom House~\cite{freedom}). Since the platform does not
collect personally identifiable information, our IRB has determined that this
project does not constitute human subjects research. Regardless, we maintain
contact with any volunteers and monitor the political situations in different
regions. Some regions have been deemed too risky to operate in (\eg Iran, Syria).
In general, we aim to balance risk with potential benefits of the measurements.

\section{Localizing Censors}\label{sec:isolating}
At a high-level, our approach works as follows:
First, we use the traceroutes gathered by the ICLab platform to construct
boolean clauses such that the literals in the clauses represent ASes observed in
the traceroute. We then use the censorship measurements associated with the
corresponding traceroutes to assign truth values to the clauses. Finally, the
clauses are converted to Conjunctive Normal Form (CNF) and used as input to an
off-the-shelf SAT solver. The process is repeated for each type of censorship
measurement (\ie DNS injection, HTTP tampering, and blockpage detection) and
various time slices (\ie for all measurements performed during the same day,
week, and month). 
Next, we analyze the solutions returned by the SAT solver for each CNF. In
cases where there are multiple solutions -- \ie multiple truth assignments for a
given CNF formulation -- we return all literals (ASes) having \emph{True}
assignments as potential censors. In cases where there is a single solution, we
return all literals (ASes) having \emph{True} assignments as censors.
Finally, we characterize censorship leakage by identifying ASes that
observe censorship only when they transit through censoring ASes.

\subsection{Constructing a SATisfiliability problem}
Each record in the ICLab dataset contains: (1) the vantage point AS, (2) the URL
being tested, (3) the anomaly being tested (and whether it was detected or not),
(4) three traceroutes between the vantage point and the URL at the time of
testing, and (5) the time at which the test was performed. We use each of these
records to create boolean satisfiability problems as follows: 

\myparab{Clause formulation.} First, we use historical IP-to-AS
mapping from CAIDA \cite{ip-as-mapping} to convert the IP-level traceroutes to
AS-level paths. Next, we eliminate cases with inconclusive paths -- \ie cases
where one of the following situations occurred: (1) IP-to-AS mapping was not
possible for any of the IPs observed in the traceroute, (2) traceroutes were not
possible due to errors, (3) AS-inference was not possible due to non-responsive
hops and different ASes observed in the previous and subsequent responsive hop,
and (4) there was more than one AS-level path obtained after conversion of the
three traceroutes. Each of the remaining AS-level paths forms a clause in our
SAT formulation, with each observed AS acting as a literal. The truth value
attached to the clause is \emph{True} if the measurement detected its
corresponding anomaly, and \emph{False} otherwise. For example, if the
AS-level path $X \rightarrow Y \rightarrow Z$ observed DNS censorship, it is
represented by the clause $(X_{DNS} \lor Y_{DNS} \lor Z_{DNS}) = T$. 

\myparab{Time- and URL-based splitting.} Our formulation accounts for the fact
that censorship policies and techniques may change over time  (\eg Iran is known
to increase censorship during political events such as elections
\cite{Aryan-FOCI13}). Not doing so introduces the possibility of generating an
unsolvable CNF in the event of a policy change -- \eg the measurement $(X_{DNS}
\lor Y_{DNS} \lor Z_{DNS}) = T$ is observed on Day 1 and $(X_{DNS} \lor Y_{DNS}
\lor Z_{DNS}) = F$ on Day 2. We address this problem by creating CNFs at four
time granularities -- days, weeks, months, and year. Additionally, since not all
URLs being tested are subject to censorship, we further restrict the CNFs to
only include clauses containing measurements to a single URL. Therefore, we
generate one CNF per URL per time granularity (day, week, month, and year).
Finally, each CNF is solved using and off-the-shelf SAT solver.

\subsection{Analyzing SAT solutions}
Given a CNF, a SAT solver may return no solution, a single solution, or
multiple solutions. When no solution is returned, there is no possible truth
assignment to ASes that can satisfy the input CNF. These scenarios may arise due
to (1) noise in the ICLab measurements -- \ie incorrect anomaly detection or
path inference or (2) changing censorship policies within the specified time
granularity. A single solution implies the presence of exactly one satisfying
truth assignment to ASes. This ideal scenario allows us to exactly identify ASes
that are responsible for generating the measured censorship related anomalies
(\ie the ASes that are assigned a \emph{True} value in the solution). We label
these ASes as \emph{censoring ASes}. Finally, when the CNF does not contain
enough clauses to generate a single solution, multiple satisfying assignments
may be possible. When this situation arises, we consider every AS as a
\emph{potential censor} unless the literal associated with it is assigned a
\emph{False} value in all returned solutions. 

\subsection{Identifying censorship leakage}\label{sec:leakage}
In order to prevent leakage of censorship (\ie where regional censorship
policies impact users outside the region), censorship policies need to
be implemented in ASes that are either stubs or provide transit services
only for ASes within the region. To uncover instances of censorship leakage, we
use the following approach: First, we only consider all AS-level paths used in
CNFs that return exactly one solution. Next, we identify ASes that (1) are
assigned a \emph{False} truth value in the returned solution, (2) are located
upstream from the identified censors (\ie closer to the vantage point being used
by ICLab), and (3) are located in a different country from the censoring
ASes in the CNF. We label these ASes as victims of censorship leakage due to
their inheritance of censorship from censoring ASes in other countries. This
inheritance occurs due to their traffic transiting through censoring ASes.

\section{Experimental Results}\label{sec:censorship}
We focus on measuring (1) how often our approach is able to generate solvable
SAT instances, (2) the amount of path churn observed and its impact on our SAT
instance solvability, and (3) the ASes responsible for implementing and leaking
censorship.

\begin{figure}[t]
\begin{subfigure}[b]{0.495\textwidth}
\scalebox{1}{
\centering
\includegraphics[trim=0cm 0cm 0cm 0cm, clip=true,width=.8\textwidth]
{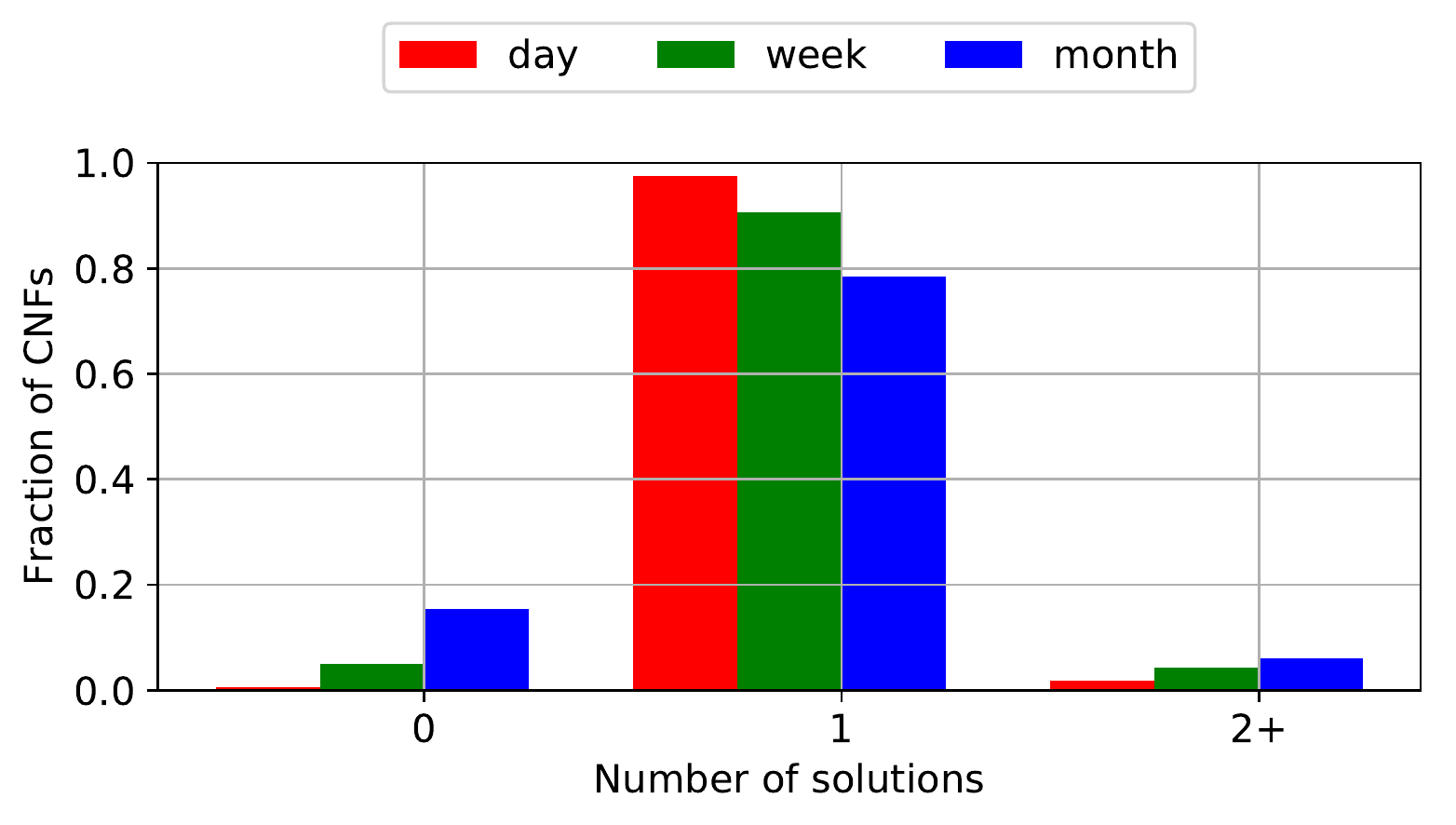}}
\caption{By CNF granularity.}
\label{fig:solutions-granularity}
\end{subfigure}

\begin{subfigure}[b]{0.495\textwidth}
\scalebox{1}{
\centering
\includegraphics[trim=0cm 0cm 0cm 0cm, clip=true,width=.8\textwidth]
{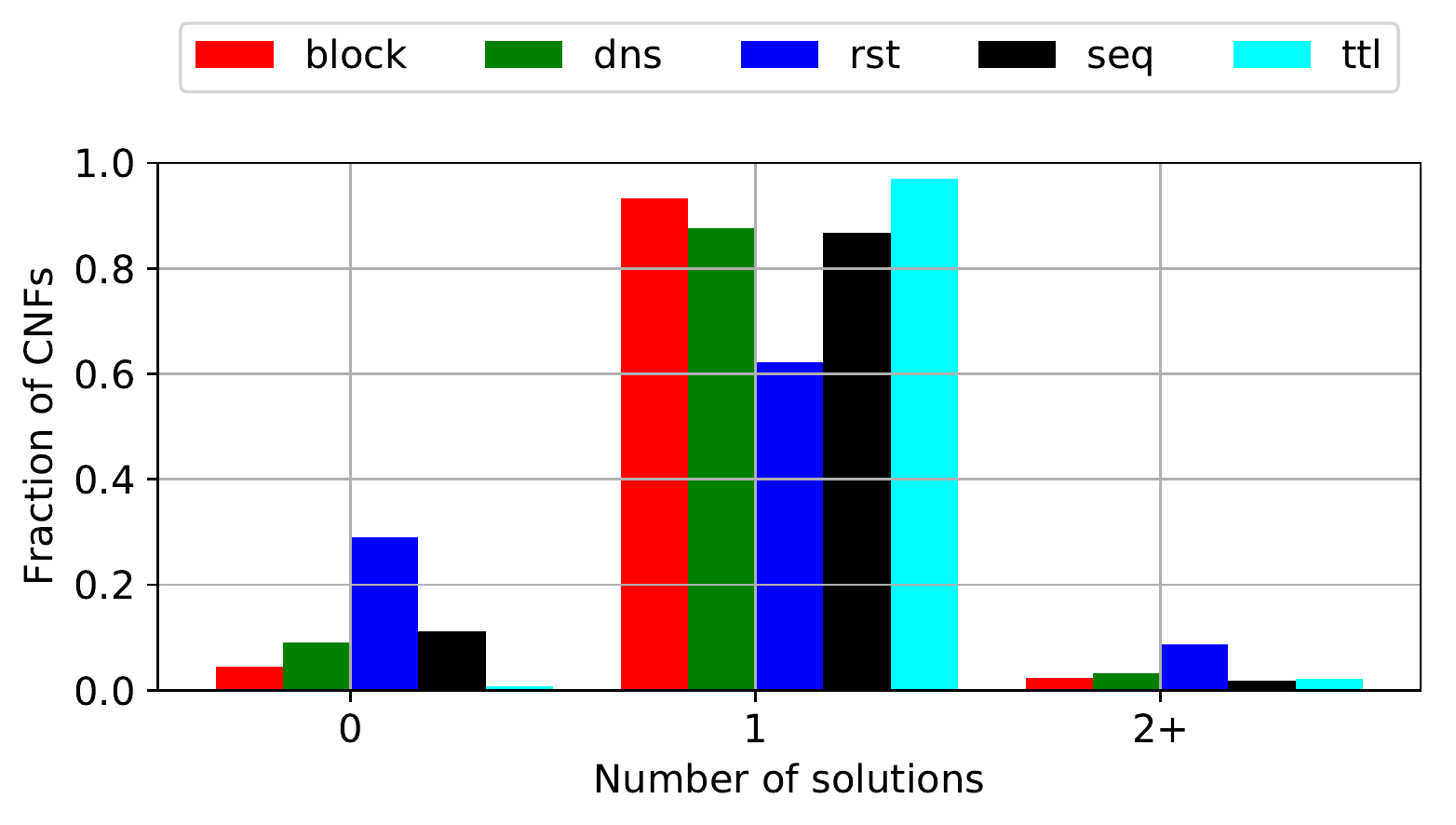}}
\caption{By anomaly type.}
\label{fig:solutions-type}
\end{subfigure}
\caption{Number of solutions found for constructed CNFs when split by CNF
granularity and anomaly.}
\vspace{-.1in}
\label{fig:solutions}
\end{figure}

\myparab{Satisfiability of generated CNFs.} As discussed earlier, the CNFs generated
by our approach may return (1) no solutions -- indicative of changing censorship
policies or noise in ICLab measurements, (2) exactly one solution -- the ideal
scenario, or (3) many solutions -- indicative of insufficient number of
measurements through diverse paths. In order to understand which scenario occurs
most frequently, we analyze the results returned by our SAT solver for CNFs of
different time granularities and anomaly types. We find that on average, nearly
92\% of our CNFs return exactly one solution and less than 6\% of our CNFs
return no solution. This indicates high fidelity in our underlying data and
highlights our ability to exactly identify censoring ASes. 

Our results, when considering CNFs generated for different time granularities
and anomalies, are illustrated in \Cref{fig:solutions}.
\Cref{fig:solutions-granularity} shows that as our CNF granularity becomes
coarser, its solvability reduces. This is expected since (1) censorship policies
are more likely to change and (2) we are more likely to include a noisy
measurement in our CNF form when considering larger time periods. \Cref{fig:solutions-type}
shows that nearly 30\% of the CNFs generated to identify ASes performing RST
injection are unsolvable. This is indicative of low fidelity RST injection
measurements from the ICLab platform, due to the difficulty of differentiating
between organic and injected RST packets. 

\begin{figure}[ht]
\scalebox{.9}{
\centering
\includegraphics[trim=0cm 0cm 0cm 1.1cm, clip=true,width=.45\textwidth]
{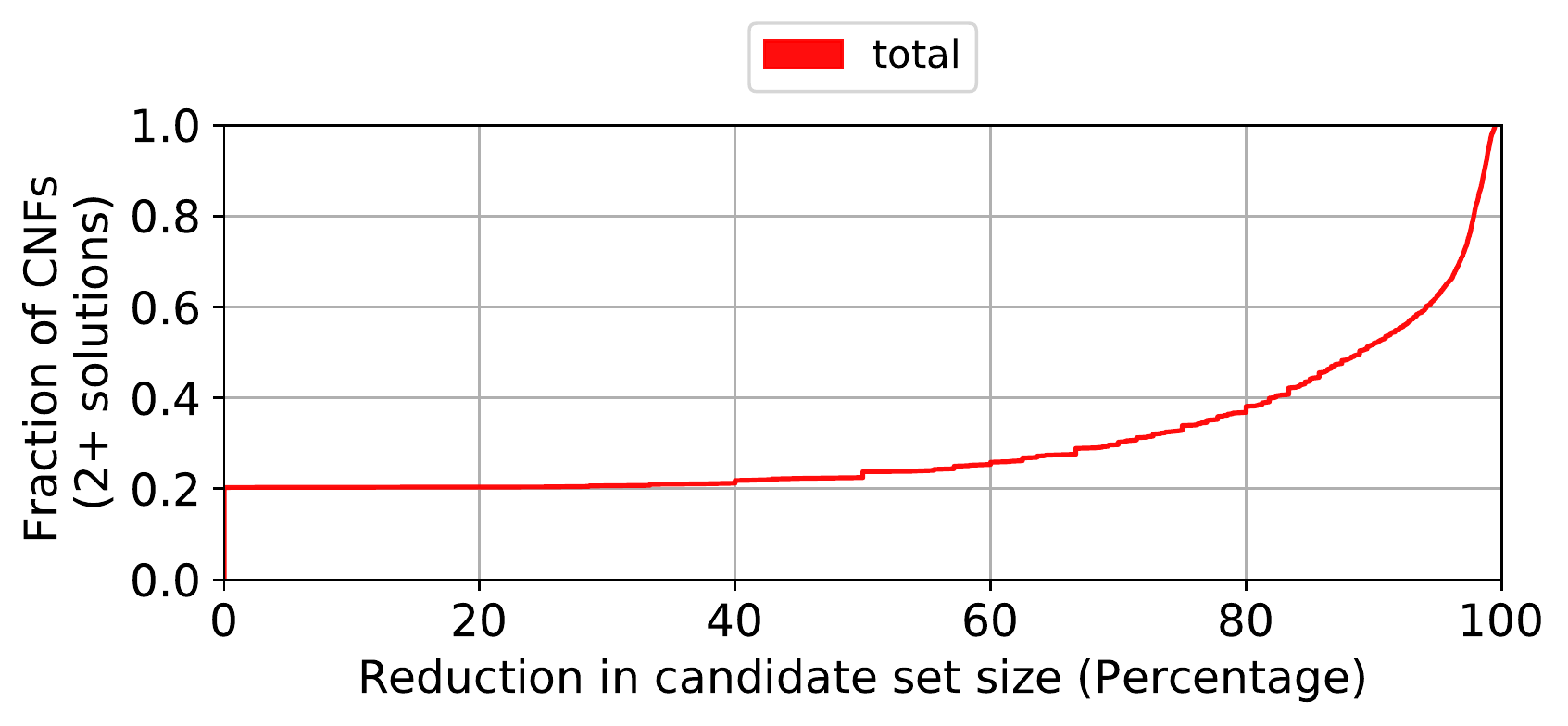}}
\vspace{-.1in}
\caption{CDF of reduction in number of potential censors in CNFs with 2+
solutions.}
\label{fig:solutions-reduction}
\end{figure}

Even if the CNFs generated by our approach yield more than one solution, they
can be useful to identify ASes that \emph{could not} have been responsible for
implementing censorship -- \ie ASes that were assigned a \emph{False} truth
value in every solution. We further investigate the ~3\% of scenarios where
constructed CNFs yield more than one solution to identify the impact of this
reduction. We find that in 20\% of such cases, the solutions satisfying the CNF
do not allow for any elimination of ASes as censors -- \ie we are unable to
narrow down the set of possible censors by eliminating definite non-censors.
However, on average, 95.2\% of all ASes in a CNF are identified as definite
non-censors, leaving only 4.8\% of the observed ASes as potential censoring
ASes. \Cref{fig:solutions-reduction} shows that 50\% of all generated CNFs with
multiple solutions have nearly 90\% of their ASes eliminated as potential
censors.

\begin{figure}[ht]
\scalebox{.9}{
\centering
\includegraphics[trim=0cm 0cm 0cm 0cm, clip=true,width=.45\textwidth]
{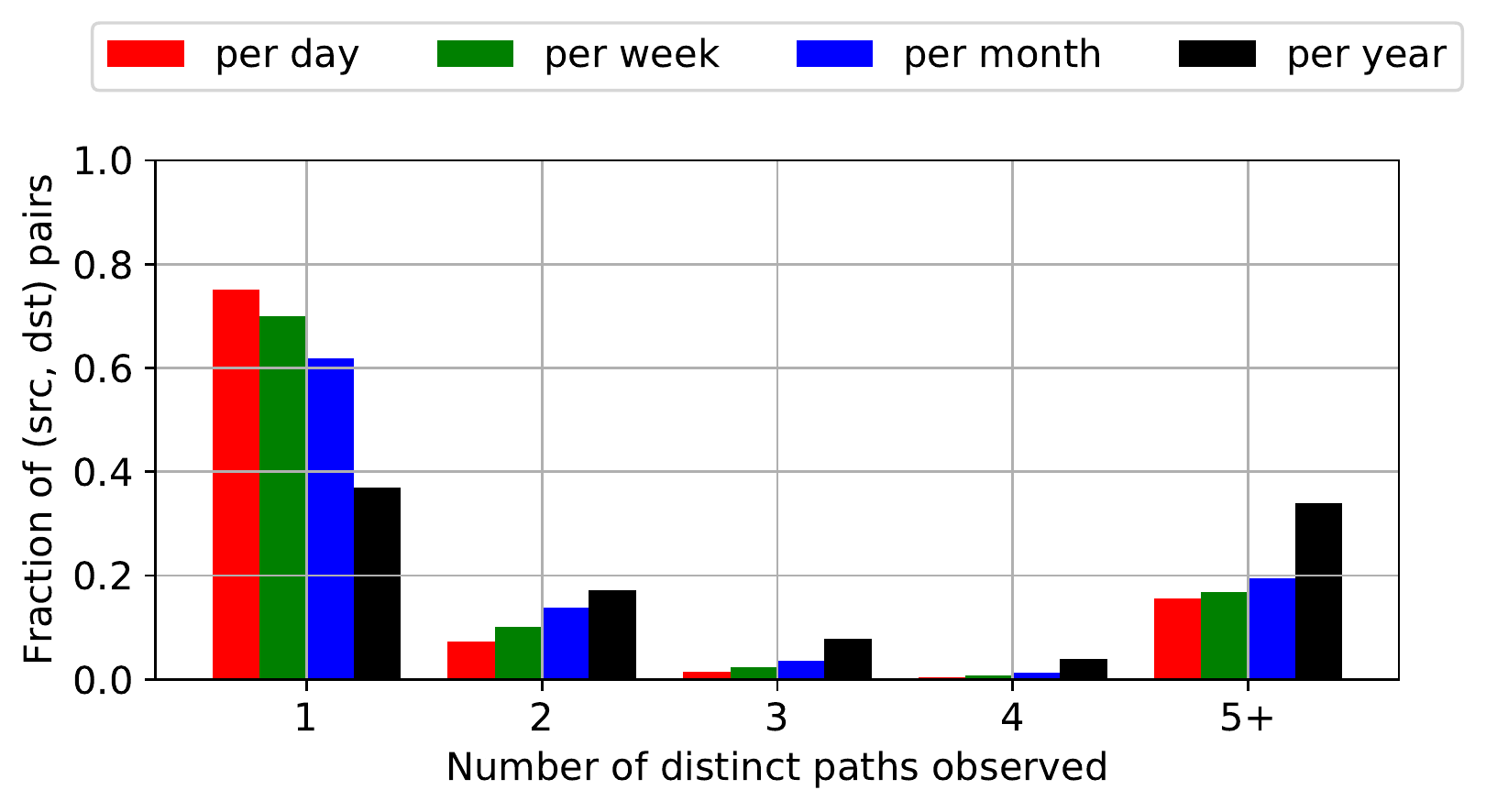}}
\caption{Number of distinct paths observed between a source and destination AS
over varying time periods.}
\vspace{-.1in}
\label{fig:churn-basic}
\end{figure}

\begin{figure}[t]
\scalebox{.9}{
\centering
\includegraphics[trim=0cm 0cm 0cm 0cm, clip=true,width=.45\textwidth]
{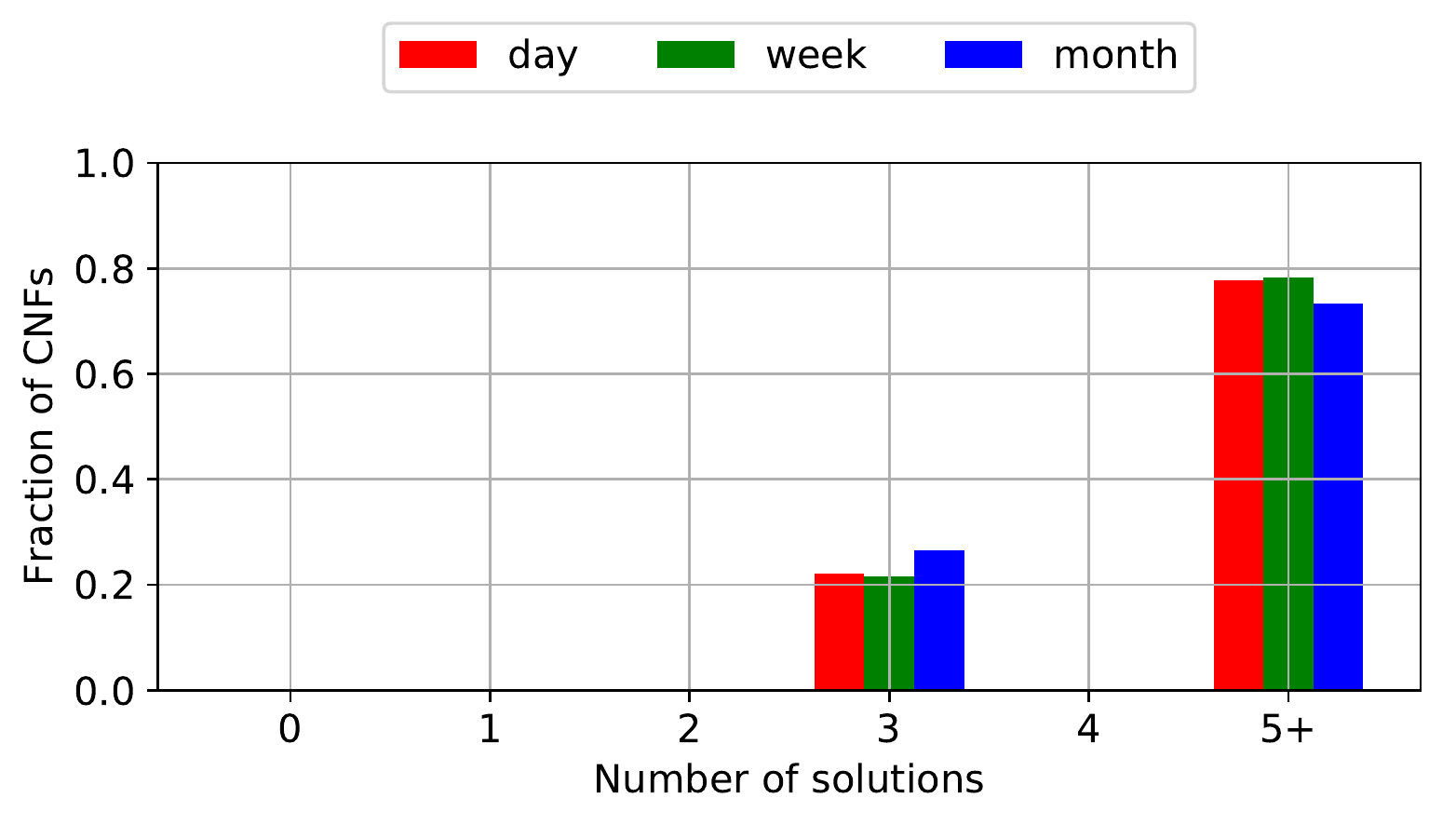}}
\caption{Number of solutions returned by CNFs in the absence of network-level
path churn.}
\vspace{-.1in}
\label{fig:churnless-solvability}
\end{figure}

\myparab{Impact of path churn.} Now, we measure the amount of network-level path
churn observed over the course of our measurements and consider the impact it
has on the usability of our approach. In order to measure the amount of
network-level path churn, we measure the number of distinct network-level paths
observed between each source (ICLab vantage point) and destination URL for each
day, week, month, and year. \Cref{fig:churn-basic} shows the fraction of these
(source, destination) pairs that observe path changes over varying time periods.
We find that nearly 25\% of all pairs observe path changes within a single day.
This fraction increases to 30\%, 38\%, and 67\% when considering periods of one
week, month, and year, respectively. Over the period of one year, 35\% of the
measured pairs recorded at least five distinct network-level paths. Using
CAIDA's AS classification database \cite{CAIDA-ASClass}, we found no significant
differences in the amount of churn observed when considering specific classes of
destination ASes (content, enterprise, or transit AS). 

To understand the impact of network-level path churn on the effectiveness of our
approach, we analyze the solvability of CNFs constructed in the absence of path
churn. We eliminate the impact of path churn by only considering the
measurements using the first observed distinct path between a source and
destination in each CNF. \Cref{fig:churnless-solvability} illustrates the
number of solutions returned by such CNFs. We see that nearly 80\% of all CNFs
return five or more solutions (compared to < 1\% in the case of CNFs that
include multiple distinct paths). 
We find that although less than 25\% of all paths are impacted by path churn
each day, the impact of these path changes on the solvability of the constructed
CNFs is significant.

\myparab{Uncovering censors and censorship leakage.} We now analyze the
censoring ASes identified by our approach. In total, we identify 65 censoring
ASes located in 30 different countries. The countries with the most number of
censoring ASes are reported in \Cref{tab:censors}. In our analysis we observe a
few regions implementing a wide array of censorship approaches. In particular,
we find that censors in China and Cyprus implement all measured forms of
censorship. Further, the ASes AS1299 (Telianet, Sweden), AS59564 (UNIT-IS,
Ukraine), and AS8966 (Etisalat, UAE) are all found to implement at least four of
the five measured censorship approaches. Using the McAfee URL categorization
database \cite{McAfee-URLs}, we find that URLs that are most commonly censored
fall in the Online Shopping and Classifieds categories. Further analysis reveals
that most ASes perform censorship exclusively on few categories of sites, with
the exception of ASes in Cyprus which censor content across many different
categories. Interestingly, we also identify several ASes (located in Ireland,
Spain, and the United Kingdom) which exclusively censor URLs associated with
popular ad vendors. 

\begin{table}[ht]
\scalebox{.9}{
\centering
\small
\begin{tab}{l|p{1.45in}|p{.9in}}
\textbf{Region} & \textbf{Censoring ASes} & \textbf{Anomalies}
\\\midrule
China & AS4132, AS4812, AS4837, AS17621, AS37963, AS58461 & All\\
United Kingdom & AS5413, AS8928, AS9009, AS20860, AS35017, AS42831 & Block, TTL\\
Singapore & AS4657, AS7473, AS17547, AS38001 & SEQ, TTL\\
Poland & AS20853, AS31621, AS42656 & Block, DNS, SEQ\\
Cyprus & AS8544, AS35432, AS197648 & All\\
\end{tab}}
\caption{Regions with most number of censoring ASes.}
\vspace{-.20in}
\label{tab:censors}
\end{table}

\begin{figure}[t]
\centering
\includegraphics[trim=0cm 7cm 0cm 2cm, clip=true,width=.495\textwidth]
{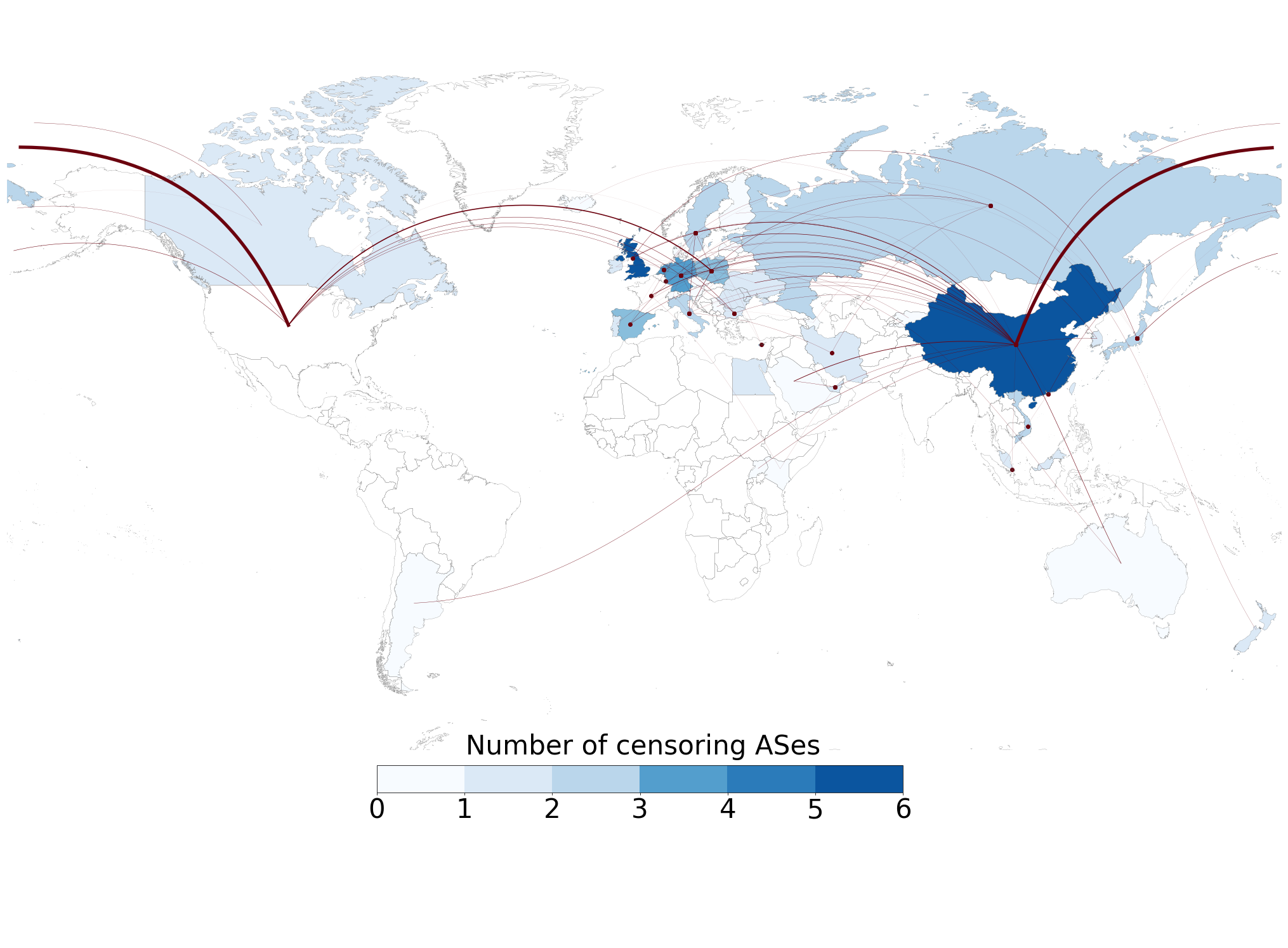}
\caption{Flow of censorship. Darker countries contain more censoring ASes and
thicker lines indicate more leakage. Sources of leakage are marked
with a point.}
\vspace{-.1in}
\label{fig:leakage-flows}
\end{figure}

To identify which of these 65 censoring ASes leak censorship, we record the
regions for which they provide upstream transit (\S \ref{sec:leakage}). We find
a total of 32 censoring ASes leak their censorship policies to other ASes. Of
these, 24 have censorship leakage extending to other countries.
\Cref{tab:censor-leakage} lists the ASes responsible for the largest number
of leaks outside their region. We find six Chinese ASes in the Top 10, with four
other ASes from Poland, Japan, Russia, and the UAE completing the list.
\Cref{fig:leakage-flows} illustrates the countries most impacted by censorship
leakage and those containing censoring ASes. We see that with the exception of
China, most other leakage is regional -- \eg European and middle-eastern censors
leak censorship mostly to other countries in the same region.

\begin{table}[ht]
\scalebox{.9}{
\centering
\small
\begin{tab}{l l p{.35in}p{.65in}}
\textbf{AS} & \textbf{Region} & \textbf{Leaks (AS)} & \textbf{Leaks (Country)}
\\\midrule
AS58461 Hangzhou-IDC & China & 49 & 21 \\
AS37963 Alibaba-CN & China & 36 & 19 \\
AS31621 QXL-NET & Poland & 28 & 13 \\
AS4812 Chinanet-SH & China & 16 & 9 \\
AS4134 Chinanet-Backbone & China & 12 & 8 \\
\end{tab}}
\caption{Censoring ASes with the largest number of censorship leaks in terms of
ASes and countries.}
\vspace{-.2in}
\label{tab:censor-leakage}
\end{table}

\section{Conclusions}\label{sec:discussion}
In this work, we leveraged boolean network tomography and censorship
measurements obtained from the ICLab censorship measurement platform to identify
the ASes responsible for inducing censorship related anomalies on the Internet.
Our results show that even in the absence of strategically selected monitors and
vantage points, exact identification of censoring ASes is possible due to
network-level path churn. Our approach uncovered 65 censoring ASes located in
(30 different countries) of which 24 were found to leak censorship into other
countries. In cases where exact identification of censors was not possible, we
were able to reduce the number of potential censoring ASes by over 95\%. 

The results obtained in this work also uncover the need to improve the fidelity
of the RST anomaly detection technique used by ICLab and traceroutes gathered by
the platform. In addition to improving the robustness of ICLab measurements (\eg
by using tools such as InTrace \cite{intrace} in conjunction with standard
traceroutes), we also plan to use our approach to extend the ICLab censorship
measurement platform in several ways. Specifically, we plan to (1) incorporate
data obtained from external performance measurement datasets (\eg data from M-Lab
\cite{MLab}) to identify ASes responsible for throttling the bandwidth made
available to specific protocols used for censorship circumvention and (2)
identify, at scale, the ASes responsible for blocking access to Tor bridges
\cite{Fifield-FOCI16}.

\bibliographystyle{ACM-Reference-Format}
\bibliography{bibliography} 

\end{document}